\documentclass[12pt]{article}
\usepackage{fleqn}

\newcommand{\be}[1]{\begin{equation} \label{#1} }
\newcommand{\eq}{\end{equation}}
\newcommand{\bea}[1]{\begin{eqnarray} \label{#1} }
\newcommand{\eea}{\end{eqnarray}}

\def\vare{\varepsilon}

\def\d{\delta}
\def\g{\gamma}

\def\l{\lambda}

\def\o{\omega}
\def\q{\theta}			
\def\s{\sigma}			

\def\D{\Delta}

\def\cf{\mathcal{F}}
\def\cg{\mathcal{G}}

\def\ci{\mathcal{I}}

\def\cl{\mathcal{L}}
\def\cm{\mathcal{M}}

\def\cs{\mathcal{S}}

\def\cx{\mathcal{X}}

\def\pa{\partial}			
\def\bpa{\bar{\partial}}	 
\def\half{\frac{1}{2}}
\def\tr{\mathrm{Tr}}			

\newcommand{\NPB}[1]{Nucl.\ Phys.\ \textbf{B#1}}
\newcommand{\PLB}[1]{Phys.\ Lett.\ \textbf{B#1}}
\newcommand{\PRD}[1]{Phys.\ Rev.\ \textbf{D#1}}
\newcommand{\xxx}[1]{\mbox{hep-th/\textbf{#1}}}

\def\MR{M.~Ro\v{c}ek {} }
\def\AGMR{A.~Giveon, \MR}

\makeatletter
\@addtoreset{equation}{section}
\@addtoreset{equation}{subsection}
\def\theequation{\ifnum\value{section}=0 \arabic{equation}\ignorespaces
\else \ifnum\value{subsection}=0 \thesection.\arabic{equation}\ignorespaces
\else \thesection.\arabic{subsection}.\arabic{equation}\ignorespaces
                       \fi
                 \fi}
\@addtoreset{table}{section}
\@addtoreset{table}{subsection}
\def\thetable{\ifnum\value{section}=0 \arabic{table}\ignorespaces
\else \ifnum\value{subsection}=0 \thesection.\arabic{table}\ignorespaces
\else \thesection.\arabic{subsection}.\arabic{table}\ignorespaces
                    \fi
              \fi}
\makeatother

\def\myint{- \half\int d^{2}\s}
\def\mint{\int d^{2}z}
\def\dil{\mathbf{\Phi}}

\def\one{\mathbf{1}}
\def\bA{\bar{A}}
\def\bJ{\bar{J}}
\newcommand{\WZW}[1]{I\Bigl[#1\Bigr]}
\def\lra{\longrightarrow}

\begin{document}

\begin{titlepage}

\begin{flushleft}
ITP-SB-95-21 \\
hep-th/9507014 \\
July 3, 1995
\end{flushleft}

\begin{center}

\vskip3em
{\bf \huge Non-Abelian Axial-Vector \\ \vskip1ex
Duality: \\ \vskip1ex a Geometric Description}

\vskip3em
{\Large\slshape Eugene Tyurin}
\vskip1em

{\slshape Institute for Theoretical Physics, SUNY at Stony Brook, \\
Stony Brook, NY 11794-3840, USA \\ \vskip1ex E-mail: \hskip1ex
gene@insti.physics.sunysb.edu }

\end{center}

\vfill

\begin{abstract}

\noindent
We give a geometric characterization of the quasi axial-vector
(Kiritsis-Obers) target space duality in the spirit of the bi-algebra
(Klim\v{c}ik-\v{S}evera) approach. We show that the $\s$-models
constructed by taking quotients have non-abelian chiral
currents that obey
 ``non-commutative conservation laws'' and provide the
criterion for a $\s$-model to have a dual using the axial-vector
procedure.

\end{abstract}

\vfill\hfill {\footnotesize Typeset by \LaTeXe}

\end{titlepage}

\section{Introduction}\label{intro}

The field of non-abelian target space duality is the next, more
complicated step towards a possible understanding of the global
structure of the space of string vacua and, in particular,
compactification and cosmological issues.

The original form of non-abelian duality \cite{AGMR,que} started with a
$\s$-model that was acted on by a group of non-abelian isometries; this
led to the puzzle of the ``vanishing isometries'': the inability
to perform an inverse duality transformation, because the model obtained
did not have the isometries of the original $\s$-model.

Besides \cite{AGMR,que}, there were several more attempts to develop a
different approach to non-abelian duality: works by Sfetsos \cite{gwzw},
\'{A}lvarez, \'{A}lvarez-Gaum\'{e} and Lozano \cite{T,Y}, Kiritsis and
Obers \cite{axvect}, and, the most encouraging one, by Klim\v{c}ik and
\v{S}evera \cite{sever}.

In the abelian case, there are two\footnote{There is
also a Hamiltonian approach to the T-duality (for a review see e.g.
\cite{T}), but it is beyond the scope of this paper.} equivalent ways
to obtain dual models: the method of Lagrange
multipliers \cite{bush} and the method of axial-vector quotients
\cite{dqc}.  It turns out that these two methods yield mutually
incompatible results when applied to the $\s$-models with non-abelian
isometries.

Recently, Klim\v{c}ik and \v{S}evera \cite{sever} suggested
a way to give up
one of the central concepts of T-duality: the need for an isometry of
the target space manifold.  Their \emph{generalized non-abelian duality}
(GNAD) is built around the \emph{bi-algebra of quasi-isometries} (as
opposed to the usual path-integral manipulations) and it includes the
``traditional''\cite{AGMR,que} non-abelian duality as a special case.

In this letter we analyze non-abelian axial-vector duality (NAAVD)
\cite{axvect} and find that it produces a family (parametrized
by the group of external automorphisms of the isometry group) of dual
models that admit ``non-commutative conservation laws'' \cite{sever}.
They do not form a bi-algebra, but by
using them, we have been able to give the Kiritsis-Obers duality a
geometric characterization: a condition that has to be satisfied by a
$\s$-model in order to have a dual model.  Such condition
replaces the need for the existence of Killing vectors; and it is
different from the one needed for the Klim\v{c}ik-\v{S}evera duality.

It seems highly unlikely that these two approaches can be reconciled,
though this has not been rigorously proven yet.  This suggests that
the unified picture of target space dualities may be even more complex
than conjectured in \cite{sever}.

The outline of the paper is as follows.  First, we briefly describe
abelian axial-vector duality \cite{dqc,oddz}.  Next, we summarize
the core notions of the GNAD \cite{sever}.  In Section \ref{QNAD} we
study the axial-vector duality of non-abelian isometries \cite{axvect}
and show how it relates to the ideas of GNAD. The conclusions and the
discussion are presented in the Section \ref{concl}.

\section{Abelian Duality and Quotients}\label{abq}

We start with the most general $\s$-model action that admits abelian
(anti)-chiral currents:
\bea{D2}
\cs_{D+2}\Bigl[\q_i,x\Bigr] &=& \myint \Biggl[ \: \pa\q_1\bpa\q_1 +
\pa\q_2\bpa\q_2 + 2B(x)\pa\q_2\bpa\q_1  \nonumber \\
& & + F_{i1}(x)\pa x^i\bpa\q_1 + F_{2j}(x)\pa\q_2\bpa x^j \: \Biggr]
+ \cs_\cx\Bigl[x,\dil\Bigr] \\ \label{cscx}
\cs_\cx &=& \myint \Biggl[ \: F_{ij}(x)\pa x^i\bpa x^j -
\frac{\sqrt{\g}}{4}\mathbf{R}^{(2)}\dil(x) \: \Biggr], 
\eea
where the conserved currents are given by
\be{abcurts}
J = \pa\q_1 + B\pa\q_2 + \half F_{i1}\pa x^i \qquad
\bJ = \bpa\q_2 + B\bpa\q_1 + \half F_{2j}\bpa x^j
\eq
We can take a quotient with respect to the axial-vector symmetry of
(\ref{D2}) using the minimal coupling ($\vare = \pm 1$)
$$
\pa_\pm\q_1 \longrightarrow \pa_\pm\q_1 + A_\pm \qquad
\pa_\pm\q_2 \longrightarrow \pa_\pm\q_2 + \vare A_\pm
$$
and the gauge choice $\q_2=0$, combined with the addition of the
following piece:
\be{extra}
\D\cs_{D+2} =  \myint \:\vare \Biggl[ \pa\q_1\bpa\q_2 -
\pa\q_2\bpa\q_1 \Biggr],
\eq
which, upon gauging, becomes
\be{extra2}
\myint \Biggl[ \bA \pa\q_1 -A\bpa\q_1 \Biggr];
\eq
this ensures the invariance of the action (\ref{D2}) under finite
gauge transformations.  After the gauge fixing, the quotient action has
the following form (we have dropped the subscript of $\q_1$):
\bea{qact}
\cs_q\Bigl[\vare,\q,x\Bigr] &=& \myint \Biggl[ \:
\frac{1-\vare B}{1+\vare B}\pa\q\bpa\q -
\frac{\vare F_{2j}}{1+\vare B}\pa\q\bpa x^j \nonumber \\
& & + \frac{F_{i1}}{1+\vare B}\pa x^i\bpa\q +
\Bigl( F_{ij} - \frac{\vare}{2}\frac{F_{i1}F_{2j}}{1+\vare B}
\Bigr)\pa x^i\bpa x^j
\\ \nonumber & & - \frac{\sqrt{\g}}{4}\mathbf{R}^{(2)}\Bigl(\dil +
\ln\det (1 + \vare B) \Bigr) \: \Biggr],
\eea
where the correction to the dilaton field $\dil(x)$ was acquired
while integrating out the gauge fields $A_\pm$ by their equations of
motion and it is equal to the corresponding Jacobian.  It is easy to
see that the two quotients corresponding to $\vare = 1$ and
$\vare = -1$ are T-dual to each other.  The above result can be
straightforwardly generalized for the case of $d$ abelian isometries
yielding a family of mutually dual models \cite{oddz}.

As we have just seen, in the case of abelian isometries, the notions of
duality and quotient are closely related: a dual pair of $\s$-models can
be constructed from a higher-dimensional $\s$-model; and, conversely,
any two axial-vector
quotients are related by a suitable duality transformation.

In this article we intend to give a geometric characterization for the
non-abelian case.  We will show how, given an abstract $\s$-model, one
can determine if there exists a dual to it that can be obtained by the
quotient procedure.

\section{Generalized Non-Abelian Duality}

In this section we introduce the formalism of generalized
non-abelian duality (GNAD) proposed by Klim\v{c}ik and \v{S}evera
\cite{sever}.  The ``traditional'' non-abelian target space duality
\cite{AGMR,que} can be seen as a particular case of GNAD when the
``dual group'' (cf. (\ref{dgginv}),(\ref{cm})) happens to be abelian.

The central idea of \cite{sever} was to abandon the requirement for
the $\s$-model to have a set of isometries (and corresponding to
them conserved currents) in favor of so-called ``non-commutative
conservation laws''.  This is a weaker property that can be
found in the dual theories.

Suppose we are given some $\s$-model (\ref{cscx}) defined on a target
manifold $\cm$ and we vary the coordinates on $\cm$ using the action of
some Lie group $\cg$: $\d x^i = \vare^a {v^i}_a$, where
 ${v^i}_a( x^i )$ is a (left) invariant frame field on
the group manifold.
We may define\footnote{As we are going to see in the
next section, non-abelian axial-vector duality provides us with the
$\s$-models that violate (\ref{curr}).  This makes them interestingly
distinct from the case considered by Klim\v{c}ik and \v{S}evera.}
the Lie derivatives of the background fields and the currents equal to
\be{lieder}
\cl_{v_a}(F_{ij}) =F_{ij,k}v^k_a + F_{kj}v^k_{a,i} +
F_{ik}v^k_{a,j}, \qquad \cl_{v_a}(\dil(x^i))=\dil_{,k}v^k_a
\eq
\be{curr}
\bJ_a = v_a^iF_{ij}\bpa x^j, \qquad J_a = v_a^iF_{ji}\pa x^j
\eq
Then the variation of the action is
\be{variation}
\cs_\s\Bigl[x^i + \vare^a v^i_a\Bigr] - \cs_\s\Bigl[x^i\Bigr] =
\mint \Bigl[ \: \vare^a \cl_{v_a}(L) + \pa\vare^a \bJ_a+
\bpa\vare^a J_a \: \Bigr],
\eq
where $\cl_{v_a}(L)$ schematically denotes the terms coming from
(\ref{lieder}).  Note that $\cg$ needs not to be an isometry group
of $\cm$, since that would amount to the much stricter requirement
$\cl_{v_a}(L)=0$ than the one we are going to impose.

We will say that the given model admits ``non-commutative conservation
laws'' if the currents (\ref{curr}) can be locally represented
(on-shell) as
\be{dgginv}
\bJ_a d\bar{z} - J_a dz =\tr(\:\tilde{T}_a \tilde{g}^{-1}d\tilde{g}\:),
\eq
where $\tilde{g}\in\tilde{\cg}$ is the ``dual quasi-isometry group'' or,
equivalently, that they obey the Maurer-Cartan equation\footnote{Note
that the usual conservation law is simply $\pa\bJ_a+\bpa J_a = 0$.}.
\be{cm}
\pa\bJ_a+\bpa J_a + {\tilde{f}_{a}}^{ bc}J_b\bJ_c = 0,
\eq
Obviously, the Lie derivative terms of (\ref{variation}) should then
satisfy the condition
\be{cm2}
\cl_{v_a}(L) + {\tilde{f}_{a}}^{bc}J_b\bJ_c = 0,
\eq
or, after substituting the (\ref{curr}), equation (\ref{cm2yes})
turns into the condition on the target $\cm$:
\be{cm2yes}
\cl_{v_a}(F_{ij}) + \tilde{f}_a^{bc} v_b^k F_{ik} v_c^l F_{lj} = 0
\eq
Note that (\ref{cm2yes}) is the central statement of the
Klim\v{c}ik-\v{S}evera approach, because it provides us with a
geometrical, non-dynamical equation.

Conversely, if we establish the off-shell validity of (\ref{cm2yes}),
then (\ref{dgginv}) and (\ref{cm}) will follow immediately (on-shell).

It is easy to see now that the dual $\s$-model should obey the same
condition as (\ref{cm2}) but with the tilded and un-tilded variables
interchanged:
\bea{cm3}
\tilde{\bJ}_a d\bar{z} - \tilde{J}_a dz &=& \tr(\:T_a g^{-1}dg\:), \qquad
g \in\cg \\ \label{cm4}
\cl_{\tilde{v}_a}(\tilde{L}) + {f_{a}}^{bc}\tilde{J}_b\tilde{\bJ}_c &=& 0
\eea

{}From the integrability condition on the Lie derivative (\ref{cm2yes}),
one can easily check that the original and dual structure constants $f$
and $\tilde{f}$ should obey the relation
\be{bi}
f_{dc}^a \tilde{f}_a^{rs} = \tilde{f}_c^{as} f_{da}^r +
\tilde{f}_c^{ra} f_{da}^s - \tilde{f}_d^{as} f_{ca}^r -
\tilde{f}_d^{ra} f_{ca}^s,
\eq
which is known in mathematics to be the relation for the structure
constants of the Lie bi-algebra $(\cg,\tilde\cg)$ \cite{sever}.

\section{Non-Abelian Duality and Quotients}\label{QNAD}

This time we consider the most general action that has
\emph{non-abelian} $\cg_L \otimes \cg_R$ (anti)chiral symmetry
\cite{axvect}:
\be{axxx}
\hspace*{-2em}\cs_{D+2d}\Bigl[h_i,x\Bigr] = \WZW{h_1} + \WZW{h_2} +
\mint\tr\Bigl[\:\bJ_2 B(x) J_1 + \bJ_2\cf^2 + \bar{\cf}^1 J_1\:\Bigr] +
\cs_\cx,
\eq
where $\cs_\cx$ is the same as in (\ref{cscx}) and
the normalization of the
WZW actions\footnote{It can be shown that if one tries to use
the Principal Chiral Model actions instead of the WZW actions in
(\ref{axxx}), the resulting quotient $\s$-models do not reduce correctly
to the abelian case (\ref{qact}).}
is fixed by the Polyakov-Wiegmann formula
\be{pw}
\WZW{h_1 h_2} = \WZW{h_1} + \WZW{h_2} + \mint\tr\Bigl[\bJ_2 J_1\Bigr].
\eq
We have defined
\bea{def}
J_i &=& h_i^{-1}\pa h_i \qquad \bJ_i = \bpa h_i h_i^{-1} \\
\tr(\bar{\cf}^1 T_a) &=& \bpa x^i F^1_{ia}(x) \qquad
\tr(\cf^2 T_a) = F^2_{ai}(x)\pa x^i
\eea
and omitted the unnecessary factors in front of the integrals.  The
resulting chiral currents can be found to be
\be{jrjl}
J_R = J_2 + \o^2\Bigl( B J_1 + \cf^2 \Bigr) \qquad
\bJ_L = \bJ_1 + \Bigl( \bJ_2 B + \bar{\cf}^1 \Bigr)\o^1,
\eq
where $\o^i_{ab} = \tr\Bigl( h_i T_a h_i^{-1} T_b \Bigr)$.

Following \cite{axvect}, we note the chiral symmetries of (\ref{axxx})
\be{chirals}
h_1 \lra u^{-1}h_1 \qquad \mbox{and} \qquad  h_2 \lra h_2 w^{-1}
\eq
and that we can
gauge them with $A=J[h_L]$ and $\bA=\bJ[h_R]$, provided the fields
transforms as
\be{gauging}
h_1\lra u^{-1}h_1, \quad h_L\lra h_L u; \qquad
h_2\lra h_2 w^{-1}, \quad h_R\lra w h_R,
\eq
where we want $u$ and $w$ to be related by an outer automorphism of
the group $\cg$: $\hat{s}[ w {}_\bullet ] = u^{-1} \hat{s}[ {}_\bullet ]$.
An outer automorphism is a transformation of a Lie group onto
itself by something that is not its own element.

The gauged action is
\bea{gauged}
\cs_{D+2d,\mbox{ gauged}} &=& \WZW{h_L h_1} + \WZW{h_2 h_R} -
\WZW{h_L \hat{s}[h_R]} + \\ \nonumber
& & \mint\tr\Bigl[\:\bJ[h_2h_R] B(x) J[h_Lh_1] \\ \nonumber
&+& \bJ[h_2h_R]\cf^2 +
\bar{\cf}^1 J[h_Lh_1]\:\Bigr] + \cs_\cx
\eea
We can parametrize the outer automorphism $\hat{s}$ in the
chosen representation by the matrix $S$:
\be{hats}
\WZW{h_L\hat{s}[h_R]} = \WZW{h_L} + \WZW{h_R} +
\mint\tr\Bigl[\bJ[h_R] S J[h_L]\Bigr],
\eq
and rewrite the (\ref{gauged}) as
\be{ggg}
\cs_{D+2d,\mbox{ gauged}} = \cs_{D+2d} +
\mint\tr\Bigl[ \bA (\o^2 B \o^1 - S) A + \bA J_R + \bJ_L A \Bigr]
\eq
The quotient action obtained as in Section \ref{abq} (in the
gauge $h_2=1$) is given by (note that
$(\bpa h h^{-1})_a = (h^{-1}\bpa h)_b\: \o^h_{ba}$):
\be{naq}
\hspace*{-2em}\cs_q\Bigl[S,h,x] = \WZW{h} + \mint\tr\Bigl[ \bar{\cf}^1
\o^h j + 
\bar{\ci} M \ci \Bigr] + \cs_\cx\Bigl[x,\dil - \ln\det M\Bigr],
\eq
where we have defined
\be{MMM}
M = \Bigl( B^{-1} S {(\o^h)}^{-1} - \one \Bigr)^{-1}, \qquad
j_{\pm} = \pa_{\pm} h h^{-1}
\eq
and
\be{II}
\bar{\ci} = \bar{j} \o^{h^{-1}} 
+ \bar{\cf}^1, \qquad
\ci = \o^h j 
+ B(x)^{-1} \cf^2
\eq

The action of the infinitesimal $g=e^{\vare^a T_a}\in\cg$
on the right,
$h\lra hg$, translates into the same perturbation as used in
(\ref{variation}), because $(\cl_x {}_\bullet \equiv
[ x , {}_\bullet ] )$
\bea{bch}
e^x e^\vare &=& \exp \Bigl( x -
\frac{\cl_x}{e^{-\cl_x}-1}\vare\Bigr),\\
h^{-1}dh &\equiv& T_a ({v^i}_a)^{-1} dx^i
= \Bigl( e^{-\cl_x} - 1 \Bigr) d =
\Bigl( e^{-\cl_x} - 1 \Bigr) \frac{1}{-\cl_x}dx \label{leftinv}
\eea
We obtain ($(\vare f)_{bc} \equiv - \vare^a f_{abc}$)
\bea{diffs}
\o^g &=& 1 - (\vare f) \\
\d j_\pm &=& \pa_\pm \vare \o^h = \o^{h^{-1}} \pa_\pm \vare \\
\d \o^h &=& - (\vare f) \o^h, \qquad
\d \o^{h^{-1}} = \o^{h^{-1}} (\vare f) \\
\d M &=& - ( 1 + M ) (\vare f) M  \\
\d \ci &=& \pa\vare - (\vare f)\o^h j, \qquad
\d \bar\ci = \bpa\vare + \bar{j}\o^{h^{-1}}(\vare f)
\eea
And the resulting variation is
\be{delta}
\d_g\cs_q\Bigl[h,x] = \mint\tr\Bigl[ \pa\vare \bJ_q +
\bpa\vare J_q \Bigr] - \mint\: \vare^a f_a^{bc}J_q^b\bJ_q^c,
\eq
where the ``non-commutative currents'' introduced in
(\ref{cm}) are explicitly given by
\be{jq}
J_q = \o^h j + M\ci \qquad \bJ_q = \bar\cf^1 + \bar\ci M.
\eq
Here we have to stop and observe two facts: 1) the above $J_{\pm q}$
are not given by (\ref{curr}); 2) (\ref{bi}) is not obeyed for
$\tilde\cg = \cg$, unless we want them to be Borelian\footnote{ We thank
Ctirad Klim\v{c}ik for this information.}.

The contradiction can be resolved if we notice that the quotient
currents (\ref{jq}) can be written as:
\bea{jq2}
J_{qa} &=& {v^i}_a \Bigl( F_{ij} + F^{WZW (-)}_{ij} \Bigr) \pa x^j \\
\bJ_{qa} &=& \bpa x^i \Bigl( F_{ij} - F^{WZW (+)}_{ij} \Bigr){v^j}_a,
\eea
where $F^{WZW (\pm)}_{ij}$ are the WZW backgrounds with different signs
of the WZ term.

Then we find
\bea{XOXO}
\cl_{v_a}(F_{ij}) &=& \tilde f_a^{bc}
\Bigl( F_{ik} - F^{WZW (+)}_{ik} \Bigr){v^k}_b
{v^l}_c \Bigl( F_{lj} + F^{WZW (-)}_{lj} \Bigr) \\
\cl_{v_a}(F^{WZW (\pm)}_{ij}) &=& 0
\eea
Equation (\ref{XOXO}) is our main result.  It is the NAAVD analog of the
Killing equation for abelian duality and of (\ref{cm2yes}) for the
Klim\v{c}ik-\v{S}evera duality.

Our integrability condition acquires an extra term on the right hand
side when compared to (\ref{bi}):
\be{nobi}
f_{dc}^a \tilde{f}_a^{rs} = \tilde{f}_c^{as} f_{da}^r +
\tilde{f}_c^{ra} f_{da}^s - \tilde{f}_d^{as} f_{ca}^r -
\tilde{f}_d^{ra} f_{ca}^s - \tilde{f}_{dc}^a \tilde{f}_a^{rs}
\eq
It comes from the part\footnote{ which is
identically zero for GNAD } of
$[ \cl_{v_c}, \cl_{v_d} ] (F_{ij})$ cubic in background fields,
where we have used the fact that
\be{wzwone}
(F + F^{WZW (-)})_{ij} - (F - F^{WZW (+)})_{ij} = \one
\eq
This term is vital for our scheme, because for $\tilde f=f$ (\ref{nobi})
becomes simply a Jacobi identity\footnote{ We have formally introduced
$\tilde f$ in (\ref{XOXO}),(\ref{nobi}) just to make a comparison with
\cite{sever}.}.

Thus we have completed the circle of transformations for the non-abelian
axial-vector duality.  We know how to construct a pair of dual models
from a base model, and how (using the equation (\ref{XOXO})) to
determine if a given $\s$-model (in the absence of obvious isometries)
can be dualized.

The one parameter family of theories (\ref{naq}) has the duality group
$Aut(\cg)$, which is the group of outer automorphisms of $\cg$, as the
automorphism parameter matrix $S$ enters into $M$.

\section{Conclusions}\label{concl}

Non-abelian axial-vector duality provides a useful tool to gain insight
into the rich geometric structure of duality which has been hidden for
some time by the simplicity of abelian systems.  Right now there are two
procedures: 1) axial-vector, which has a clear path-integral
interpretation; 2) bi-algebra, where we have a clear classification, but
no underlying path-integral mechanism.

The old $\int\tr (\l F)$ mechanism \cite{bush} turns out to be
inadequate for this more complex environment; and, should the target
space duality be a relevant symmetry of the string theory, we need to
find a unified description for it.

The other problem to be solved for Kiritsis-Obers duality is the
conformal structure of the dual models.  As we know, $U(1)$ axial-vector
quotients are identical from the CFT point of view, and one needs to
find out if this situation persists for non-abelian groups.

\section*{Acknowledgments}

It is a great pleasure to thank Martin Ro\v{c}ek for suggesting
this problem to me and for his guidance and advice.  I am also
grateful to Dileep Jatkar and Ctirad Klim\v{c}ik for stimulating
discussions.


\end{document}